\documentclass[conference]{IEEEtran}
\IEEEoverridecommandlockouts
\usepackage{cite}
\usepackage{amsmath,amssymb,amsfonts}
\usepackage{graphicx}
\usepackage[hidelinks]{hyperref}
\usepackage{url}
\usepackage{textcomp}
\usepackage{xcolor}
\usepackage{algorithm}
\usepackage[noend]{algpseudocode}
\usepackage{xspace}
\usepackage{makecell} 
\usepackage{graphicx} 
\usepackage{array} 
\usepackage{tabularx}
\usepackage{tabularray}
\usepackage[T1]{fontenc}

\renewcommand{\texttt}[1]{{\normalsize\ttfamily#1}}

\newcommand{\myparc}[1]{\smallskip\noindent\textbf{#1:}\xspace}

\usepackage[none]{hyphenat}

\def\BibTeX{{\rm B\kern-.05em{\sc i\kern-.025em b}\kern-.08em
    T\kern-.1667em\lower.7ex\hbox{E}\kern-.125emX}}
\begin{document}

\title{
A Federated Learning Platform as a Service for Advancing Stroke Management in European Clinical Centers
\thanks{This project is funded by the Horizon EU project TRUSTroke in the call HORIZON-HLTH-2022-STAYHLTH-01-two-stage under GA No. 101080564.}
}

\author{
    \IEEEauthorblockN{
        Diogo Reis Santos\IEEEauthorrefmark{1},
        Albert Sund Aillet\IEEEauthorrefmark{1}, 
        Antonio Boiano\IEEEauthorrefmark{2},
        Usevalad Milasheuski\IEEEauthorrefmark{2}\IEEEauthorrefmark{3} \\
        Lorenzo Giusti\IEEEauthorrefmark{1},
        Marco Di Gennaro\IEEEauthorrefmark{2}, 
        Sanaz Kianoush\IEEEauthorrefmark{3}, 
        Luca Barbieri\IEEEauthorrefmark{2} \\
        Monica Nicoli\IEEEauthorrefmark{2}, 
        Michele Carminati\IEEEauthorrefmark{2}, 
        Alessandro E. C. Redondi\IEEEauthorrefmark{2}, 
        Stefano Savazzi\IEEEauthorrefmark{3}, 
        Luigi Serio\IEEEauthorrefmark{1}
    }
    \IEEEauthorblockA{
        \IEEEauthorrefmark{1}CERN, Switzerland\\
        \IEEEauthorrefmark{2}DEIB, Politecnico di Milano, Milan, Italy\\
        \IEEEauthorrefmark{3}IEIIT, Consiglio Nazionale delle Ricerche (CNR), Milan, Italy
    }
}

\maketitle

\begin{abstract}
The rapid evolution of artificial intelligence (AI) technologies holds transformative potential for the healthcare sector. In critical situations requiring immediate decision-making, healthcare professionals can leverage machine learning (ML) algorithms to prioritize and optimize treatment options, thereby reducing costs and improving patient outcomes. However, the sensitive nature of healthcare data presents significant challenges in terms of privacy and data ownership, hindering data availability and the development of robust algorithms. Federated Learning (FL) addresses these challenges by enabling collaborative training of ML models without the exchange of local data. This paper introduces a novel FL platform designed to support the configuration, monitoring, and management of FL processes. This platform operates on Platform-as-a-Service (PaaS) principles and utilizes the Message Queuing Telemetry Transport (MQTT) publish-subscribe protocol. Considering the production readiness and data sensitivity inherent in clinical environments, we emphasize the security of the proposed FL architecture, addressing potential threats and proposing mitigation strategies to enhance the platform's trustworthiness. The platform has been successfully tested in various operational environments using a publicly available dataset, highlighting its benefits and confirming its efficacy.

\end{abstract}


\section{Introduction}
Stroke is the leading cause of severe disability worldwide and the second cause of death~\cite{Rajsic2019}. Global data show a prevalence of more than 12 million strokes per year, with more than 6 million being fatal. An estimated 30\% of stroke survivors are permanently disabled, resulting in approximately 110 million stroke survivors worldwide. This leads to the loss of 143 million disability-adjusted life years (DALYs) and an estimated cost of 27 billion euros for the European Union~\cite{Rajsic2019, Feigin2021}.

\begin{figure}
    \centering
    \includegraphics[width=1\linewidth]{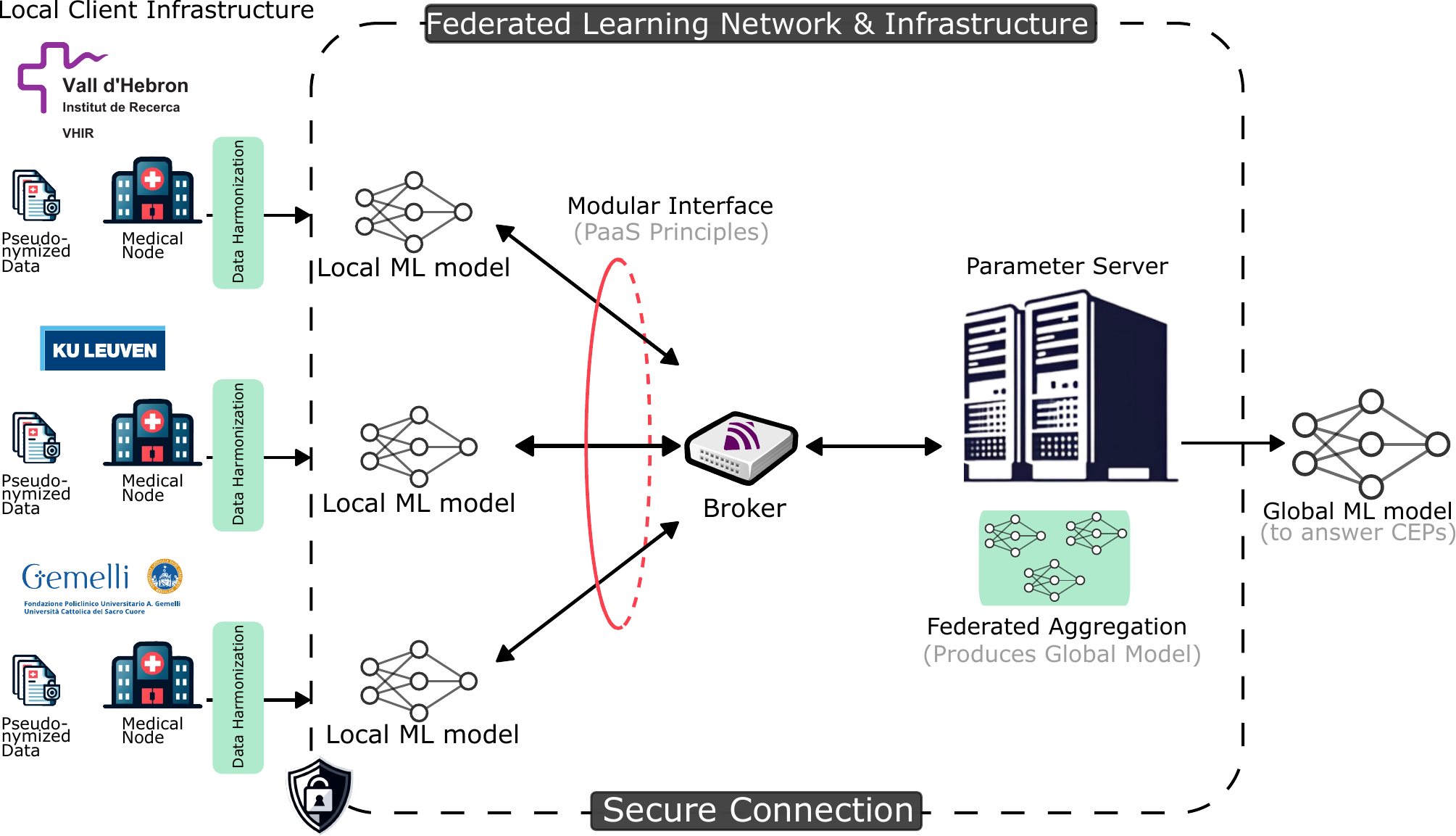}
    \caption{{TRUSTroke} scheme: data from local clinical sites is harmonized to train federated models through a Parameter Server iteratively. Model results are communicated to patients and healthcare professionals. Clinical sites are continuously involved to improve the AI models and obtain clinical evidence.}
    \label{fig:trustroke}
    \vspace{-10pt}
\end{figure}

The TRUSTroke project aims to develop a novel, trustworthy, and privacy-preserving AI platform to assist in managing both the acute and chronic phases of ischemic stroke. Leveraging clinical and patient-reported data, the project addresses five crucial clinical endpoints (CEPs): (1) clinical response to acute reperfusion treatment and stroke severity at discharge; (2) probability of early supported discharge (1 week after the event); (3) probability of poor mobility, incomplete recovery, and unfavorable long-term outcomes; (4) probability of unplanned hospital readmission (at 30 days); and (5) risk of stroke recurrence (3 and 12 months).

\begin{figure*}[ht!]
    \centering
    \includegraphics[width=\linewidth]{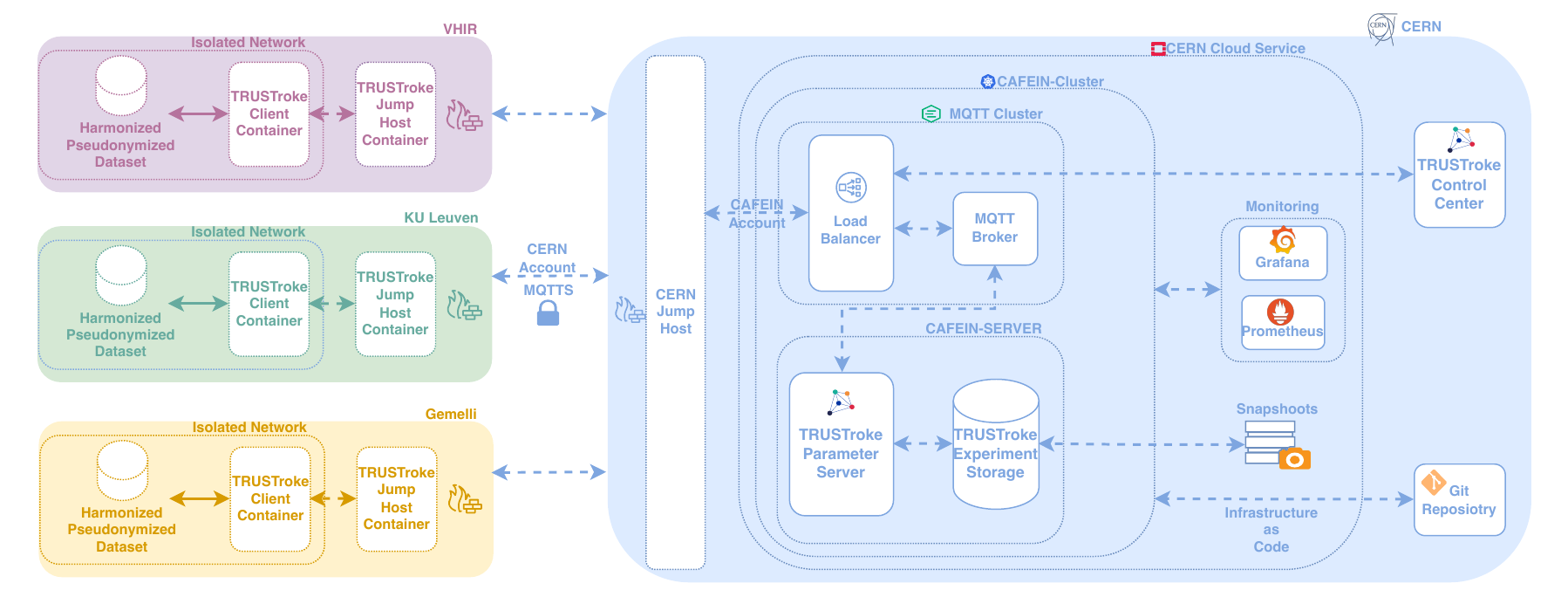}
    \vspace{-5pt}
    \caption{TRUSTroke federated learning network and infrastructure. The implementation of Client Nodes, shown on the left side, comprises two containerized applications: TRUSTroke-Jump-Host and TRUSTroke-Client. The former is responsible for communication with CERN's network and MQTT broker. The latter resides in an isolated network with data access and is responsible for training local ML models. The Broker and Parameter Server implementation is shown on the right. Based on microservices, the PS is isolated and only accessible from the MQTT broker. Experiment storage and backups are provided by the cloud infrastructure.}
    \label{fig:infra-cafein-paas}
    \vspace{-10pt}
\end{figure*}

\subsection{Federated Learning platform for stroke management}

Machine learning (ML) can play a critical role in assisting the five aforementioned CEPs with rapid, precise, and multivariate diagnosis. TRUSTroke focuses on developing a Federated Learning (FL) platform as a service targeted for clinical production environments. This privacy-preserving platform enables multiple parties to collaboratively train a model while ensuring the security of their data \cite{mcmahan2017communication, Rieke2020, Kairouz2021}. The primary concepts of the FL structure proposed in the TRUSTroke project are depicted in Fig.\ref{fig:trustroke}. Local ML models are independently trained by each medical institution using their private local data. These local models are shared with a Parameter Server (PS) that aggregates the local ML models to generate a global federated model. FL has attracted significant academic interest and is seeing growing applications, particularly in the healthcare sector and specifically for stroke management, where data privacy and security are paramount \cite{Rieke2020, Teo2024, Elhanashi2024}.

\subsection{Contributions}
A FL infrastructure, developed by CERN in collaboration with Politecnico di Milano and Consiglio Nazionale delle Ricerche and hosted at CERN, has been designed and deployed to allow multiple clinical sites to collaboratively build several trustworthy AI-based predictive models for the above-defined CEPs. This will ensure compliance with the General Data Protection Regulation (GDPR) and the European Union regulations on the storage and processing of personal data, lower hospital adoption barriers, and address the challenges identified by inspecting the EU Medical Device Regulation~\cite{Niemiec2022}, the Food and Drug Administration (FDA) repository of AI-enabled medical devices~\cite{fda2023}, and surveys on the adoption of AI in medicine~\cite{Rajpurkar2022}.

This paper introduces the proposed FL platform for TRUSTroke, considering Platform-as-a-Service (PaaS) functionalities tested and validated to support highly configurable and modular FL processes. We also assess the security threats and risks linked to each component of the proposed architecture, compiling mitigation techniques and recommendations to enhance the platform's trustworthiness. Section~\ref{sec:cafein} describes the current platform and infrastructure setup that serves as the foundation for further work. Section~\ref{sec:init} discusses the configuration and initialization of a new federated experiment. Sections~\ref{sec:process} and~\ref{sec:track} cover the orchestration and tracking of federated experiments. Section~\ref{sec:results} validates the platform's performance through real-world tests with tabular data from publicly available health records.

\section{Federated Learning Platform}
\label{sec:cafein}
CAFEIN (Computational Algorithms for Federated Environments: Integration and Networking) is a federated learning platform developed to train and deploy AI-based analysis and prediction models at CERN~\cite{cafein}. It has previously been successfully evaluated in the medical field~\cite{TEDESCHINI, roman23}. This platform serves as the background for the FL platform for the TRUSTroke project.

CAFEIN comprises four primary components: an MQTT broker, a parameter server, the client nodes, and the control center. An overview of the federated learning platform is presented in Fig.~\ref{fig:infra-cafein-paas}.

\myparc{MQTT Broker} manages message passing and communication between nodes in the federated network. It also handles the \emph{authentication} and \emph{authorization} of nodes, ensuring secure and trusted interactions within the federation. MQTT was preferred over other application protocols, such as HTTP, for its ability to manage one-to-many asynchronous communications, embedded security, and scalability features.

\myparc{Parameter Server (PS)} coordinates the training process across different nodes and is the central hub for secure aggregation. It also acts as a \emph{model server} for distributing models and \emph{experiment tracker} trace FL experiments.

\myparc{Client Nodes (CNs)} represent the medical institutions participating in the federated network. Each CN maintains access to its local data, enabling it to train and evaluate models independently and preserving data privacy.

\myparc{Control Center (CC)} acts as the administrative API and primary interface for interactions with the network after the initial setup. It facilitates the initiation and monitoring of training processes and oversees the health and status of both the PS and the CNs.

\subsection{MQTT Broker}

The MQTT Broker is deployed within a Kubernetes (K8S) cluster, utilizing CERN's Cloud Services for infrastructure \cite{cern}. EMQX MQTT broker implementation was selected based on its open-source status, native K8S support through a dedicated operator, and competitive performance metrics \cite{Longo2022}. Broker configurations are managed in a Git repository with Infrastructure as Code (IaC) principles applied to ensure that these configurations are tractable, reproducible, and security checked.

The broker manages message redirection, client authentication, and access control through an access control list (ACL). For authentication, connections are restricted to registered client identifiers (IDs) that represent individual CAFEIN accounts. By default, all subscription and publishing operations are prohibited for a particular client ID, adhering to the principle of least privilege (PoLP). Permissions for subscribing and publishing are granted to specific client IDs only as necessary. This structure allows the network manager to specify client permissions for different federations, algorithm implementations, machine learning tasks, and CEPs.

The initial release of the TRUSTroke platform will comprise three participant clients and one observational client. This observational client will not participate in the training process, but is authorized to receive the global model for evaluation. The authentication and authorization methods described above fully support these dual-client operations.

A CERN Cloud Load Balancer manages traffic ingress. The Load Balancer is accessible only from inside CERN's private network, meaning a CERN account is necessary to access the network and, thus, the MQTT broker. This setup ensures that CERN's world-leading network security advantages are integrated into the TRUSTroke platform \cite{cern}. This restriction ensures tighter security controls and limits access to critical management features to authorized personnel only.

Communication security within the network is established using TLS (Transport Layer Security). The CERN Certificate Authority (CA) generates the required certificates, ensuring that all data transmitted over the network are securely encrypted and authenticated.

\subsection{Parameter Server}

The PS is also deployed in the K8S cluster and connects to the MQTT Broker using its specific identifier (ID). By design, only one PS is allowed per federation, restricting access, improving security, and preventing misconfiguration. Connections from outside the cluster are not permitted, further reinforcing PS's isolation and security by design. Experiment artifacts, such as configuration files, logs, TensorBoards, and global models, are stored on an attached volume. These volumes are regularly snapshotted to prevent data loss and ensure data integrity.

\subsection{Client Nodes}
The software at each CN is deployed through two dockerized applications. The first, TRUSTroke-Jump-Host, creates a demilitarized zone (DMZ). This host has Internet access but does not have access to local data, ensuring a secure boundary. The second, TRUSTroke-Client, is responsible for the training process and can access local data but resides within an isolated network. It is accessible only from the TRUSTroke-Jump-Host. Connections to CERN's private network are facilitated through SSH tunneling. Firewalls are configured to allow only necessary communication through the MQTT-secured channel at a specific IP address.

\subsection{Control Center}
The CC enables users to visualize the state of the network, including connected CNs and PS, their time-stamped statuses, and any relevant diagnostic messages. CC also serves as the \emph{exclusive interface} for initiating training processes since the PS is not directly accessible.

\begin{figure}
    \centering
    \includegraphics[width=1\linewidth]
    {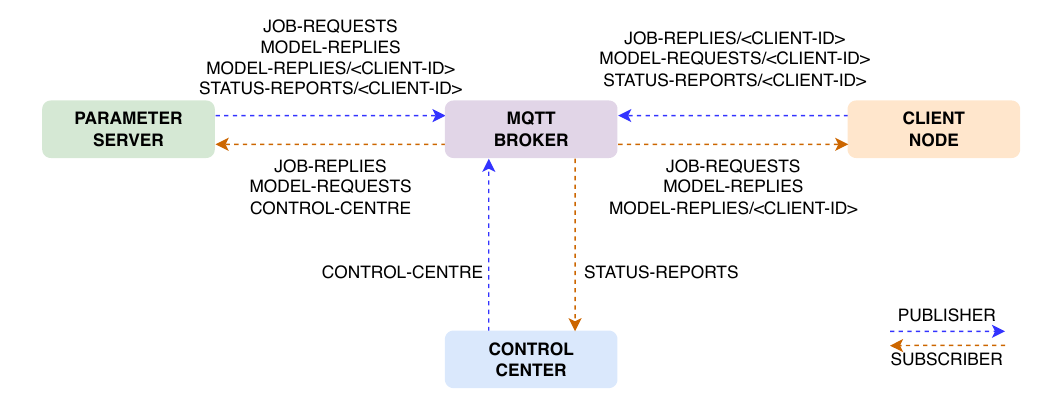}
    \vspace{-10pt}
    \caption{MQTT main PS, CNs, and CC topics. Published topics are represented by blue arrows, and subscribed topics by orange arrows.}
    \label{fig:topics}
    \vspace{-10pt}
\end{figure}

\section{FL process initialization}
\label{sec:init}
FL process initialization involves configuring various functionalities and parameters that end users can set to start the FL process. This initialization consists of: a) defining the machine learning model's architecture, b) configuring the settings for the FL experiment, and c) initiating and validating the process through the CC.

\subsection{Configuration of the machine learning model}
The TRUSTroke platform leverages KERAS version 3 to train deep learning models, supporting the three most popular back-ends: TensorFlow, PyTorch, and JAX \cite{chollet2015keras}. The platform allows for serializing any deep learning model into a JavaScript Object Notation (JSON) file. Using JSON files for model configuration offers a highly adaptive and flexible environment for users to experiment with various model configurations without requiring remote code execution or limiting them to a predefined set of models.

\subsection{FL Experiment Settings}
The Experiment Settings for the FL process can be grouped into three main areas: a) settings related to the orchestration of the federated process, b) settings for aggregation at the PS, and c) settings for local model training.

\myparc{FL Process Settings} Users can configure several aspects of the FL process. This includes specifying the number of federated rounds for the experiment, the minimum number of CN replies required for a round to be considered successful (otherwise, the round is skipped), and the timeout duration for training and evaluation on CNs (per round). After the timeout expires, CNs will stop the training or evaluation process and reply to the PS. These settings support synchronous and asynchronous FL experiments, accommodating diverse federation dynamics. In TRUSTroke, all participating CNs are considered trusted and reliable nodes, required to respond in every round. Future federation expansions could adopt different configurations. CNs can also specify if they allow TensorBoard files, containing metrics recorded during training and evaluation, to be uploaded to the PS, as these metrics could potentially leak sensitive information.

\myparc{FL Algorithm Settings} The platform supports several aggregation algorithms, including Federated Averaging (FedAvg) \cite{mcmahan2017communication}, Federated Learning with Proximal Term (FedProx) \cite{Li2018}, Federated Learning based on Dynamic Regularization (FedDyn) \cite{Acar2021}, and Stochastic Controlled Averaging for Federated Learning (SCAFFOLD) \cite{Karimireddy2019}. Based on the chosen algorithm, users can configure its specific parameters. Algorithm-specific configurations are embedded in the MQTT payload and shared with the CNs on each FL round. These parameters can be used, for example, to regulate the portion of the previous global model retained and step size of the local model aggregation process on the PS (FedAvg, FedProx) or to control the configuration of SCAFFOLD and FedDyn tools.

\myparc{Local Data Loaders} CNs must define the data loaders during the initial setup. This can consist of one or more scripts that load and preprocess local data for training or evaluation datasets. These custom scripts are necessary to accommodate the differences in data sources from various CNs, ensuring that data is harmonized for the FL process. In TRUSTroke, this is handled by a data harmonization service implemented at the four participating hospitals \cite{trustroke}.

\myparc{Local Model Training Settings} The training configuration maps to KERAS settings, allowing users to leverage its documentation and arguments. These settings align with typical machine learning setups, enabling users to define batch size, loss function, learning rate or optimizer. Additionally, custom callback mechanisms are used to tune the learning rates and configure early stopping criteria. These mechanisms are implemented on the PS and work based on the weighted mean of the post-evaluation metrics of the CNs.

\myparc{Validation} All settings are validated by an experiment schema model integrated into the platform. The experiment schema includes the specific configurations of the deployed FL algorithms. For instance, if the FedDyn tool is chosen, the user must specify the $\mu$ values \cite{Acar2021}. This validation step is initially performed by the CC when a new experiment is initiated, but it remains integral to the platform.

\subsection{Initiating a new experiment through the Control Center}
Only the CC can initiate a new experiment. After successful verification, the CC generates a unique experiment ID and sends the model configuration and experiment settings to the PS.

\begin{figure}
    \centering
    \begin{minipage}{0.48\linewidth}
        \centering
        \includegraphics[trim=1cm 1cm 0.5cm 1cm, width=\linewidth]{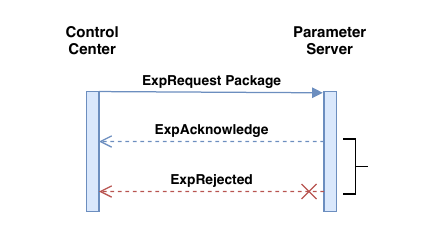}
        \label{fig:left-image}
    \end{minipage}\hfill
    \begin{minipage}{0.48\linewidth}
        \centering
        \includegraphics[trim=0.5cm 1cm 1cm 1cm, width=\linewidth]{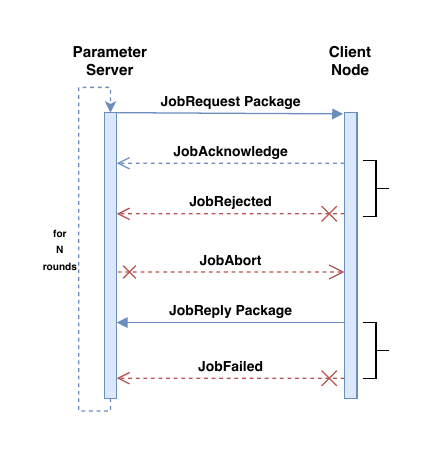}
        \label{fig:right-image}
    \end{minipage}
    \caption{Sequence diagrams for Parameter Server and Control Center interactions during FL process initialization (left). Client Nodes and Parameter Server interactions for each FL round (right).}
    \label{fig:bothimages}
    \vspace{-10pt}
\end{figure}

\section{Orchestration of a FL Experiment}
\label{sec:process}

The orchestration process involves several steps with MQTT topics, subscribers, and publishers shown in Fig. \ref{fig:topics}. Topics mentioned in the text and figures are prefixed with \texttt{CAFEIN/TRUSTroke/CEP-ID} to define a specific federation for each of the identified machine learning tasks or CEPs. Unified Modeling Language (UML) sequence diagrams in Fig. \ref{fig:bothimages} illustrate the control flow between the three federation components. Algorithm \ref{algo:1} and Algorithm \ref{algo:2} show control flow on the PS and CNs, respectively.

\myparc{Experiment Initialization} Initially, the CC publishes an \textit{Experiment Request Package} for a new FL experiment on the \texttt{control-center} topic. This package includes the model configuration and experiment settings described above, which the CC validates before sending. Upon receiving this information, the PS confirms the successful experiment initiation by replying on the \texttt{parameter-server-replies} topic. If the experiment initiation process fails, the PS replies with an \textit{Experiment Rejected} message.

\myparc{Federated Process Initialization} The PS issues a training \textit{Job Request Package} via the \texttt{job-requests} topic. This package includes the model and training settings and the current global model weights (which are initialized randomly if it's a newly initiated experiment). This request is broadcasted to all connected CNs, which, upon successfully initializing the training job, indicate their participation in the federated round by responding on individualized \texttt{job-replies/<client-id>} topics with a \textit{Job Acknowledge} message.

\myparc{Model Training and Response} CNs can perform pre- and post-evaluation of the models. Pre-evaluation measures the performance of the global model prior to fine-tuning with local data, while post-evaluation assesses it after the global model update according to the selected FL algorithm. Variations between pre- and post-evaluation may provide insights into model behavior and aid in diagnosing under-performing configurations. Toleration of failure in evaluation steps is allowed. However, post-evaluation metrics are required for the PS learning rate scheduler and early stopping mechanisms. Local model training failure is not tolerated. At the end of the training, CNs either send back the trained model with experiment artifacts through a \textit{Job Reply Package} or indicate a job execution failure with a \textit{JobFailed} message, both on their respective \texttt{job-replies/<client-id>} topics. The experiment artifacts depend on the settings but typically include the updated model weights, metrics, and TensorBoard files.

\myparc{FL Iterations and Global Model Distribution} The process iterates through several rounds, as specified by the FL experiment settings. If, for a specific round, a reduced number of \textit{JobAcknowledgments} or the number of \textit{JobFailed} messages received invalidates the minimum number of required responses, \textit{JobAbort} messages are broadcast to the CNs, and the current round is skipped at the PS. Nodes receiving \textit{JobAbort} cancel their running training or evaluation process and do not reply to the PS. At the completion of the experiment, the PS broadcasts the final global model via the \texttt{model-replies} topic, making it available to all subscribing CNs. Specific clients may request the final global model directly through the \texttt{model-requests/<client-id>} topic, and the PS responds by sending the model only to the requesting CN on the \texttt{model-replies/<client-id>} topic.

\myparc{Status Updates} Throughout their operation, both the PS and the CNs continuously publish their status updates to the \texttt{status-reports/<client-id>} topic. The MQTT broker retains these updates and can be accessed by the CC to monitor ongoing operations.

This structured description ensures a comprehensive understanding of each stage of the communication orchestration in the FL process, illustrating the critical roles of the MQTT topics and the interactions between the various components.

\begin{algorithm}[t]
\small
\caption{Parameter Server Flow}
\label{algo:1}
\begin{algorithmic}[1]
    \Statex \textbf{Input:} MQTT payload containing model configuration and experiment settings
    \State Initialize and validate the new experiment
    \State \Call{Initialize model parameters $\theta$}{}
    \State Initialize the number of federated rounds $R$
    \For{$r = 0$ to $R$}
        \State Broadcast \textit{JobRequest Package}
        \State Wait for \textit{JobAcknowledgments}
        \If{enough \textit{JobAcknowledgments} are received}
            \State Wait for \textit{JobReplies}
        \Else
            \State Broadcast \textit{JobAbort}
            \State Skip to the next round
        \EndIf
        \While{Round timeout not expired}
            \State Receive all \textit{JobReplies} or \textit{JobFails}
            \If{not enough \textit{JobReplies}}
                \State Skip to the next round
            \EndIf
        \EndWhile
        \State \Call{Aggregate Local Models}{}
    \EndFor
    \State Broadcast final global model $\theta$
\end{algorithmic}
\end{algorithm}

\begin{algorithm}[t]
\small
\caption{Client Node Flow}
\label{algo:2}
\begin{algorithmic}[1]
    \Statex \textbf{Input:} MQTT payload containing model configuration, model weights, and training settings
    \State Set Client Node to \textit{TRAINING}
    \If{pre-evaluation is set}
        \State \Call{Run Pre-Evaluation}{}
    \EndIf
    \State \Call{Run Local Training}{}
    \If{post-evaluation is set}
        \State \Call{Run Post-Evaluation}{}
    \EndIf
    \State Package experiment artifacts
    \State Send \textit{JobReply Package} to the Parameter Server
    \State Set client node to \textit{IDLE}
    \Statex
    \Procedure{HandleAnyCrash}{}
        \State Send \textit{JobFail} to the Parameter Server
        \State Set client node to \textit{IDLE}
    \EndProcedure
\end{algorithmic}
\end{algorithm}
\section{Experiment Tracking and Logging}
\label{sec:track}

Experiment artifacts for CNs and PS are stored in attached storage volumes. These artifacts include local and global models, TensorBoard files, and log files.

\begin{table*}[ht!]
\centering
\caption{Security threats associated with the proposed platform}
\label{tab:security_threats} 
\resizebox{\textwidth}{!}{  
\begin{tabular}{>{\centering\hspace{0pt}}m{0.18\linewidth}|>{\centering\hspace{0pt}}m{0.35\linewidth}|>{\centering\hspace{0pt}}m{0.11\linewidth}|>{\centering\arraybackslash\hspace{0pt}}m{0.27\linewidth}} 
\hline
\textbf{Threat} & \textbf{Description} & \textbf{Component} & \textbf{Mitigation} \\ 
\hline
Data Breach and Leakage & Unauthorized access to patient data risks privacy breaches and reputational damage & CNs & TRUSTroke-Client and Dataset are isolated from external connections \\ 
\hline
Data Interception and Tampering & Unauthorized interception or alteration of data during transmission impacts model reliability & Communication Infrastructure & MQTT with TLS encryption \\ 
\hline
Unauthorized Access & High risk of unauthorized access to client or server systems, potentially leading to data theft and system compromise & PS, CNs & Access is controlled by SSH tunneling, Kerberos, and enhanced MQTTS authentication \\ 
\hline
Denial of Service Attacks & Overloading the PS MQTT broker disrupts the training process & PS & Implementation of MQTTS reduces the risk of DoS \\ 
\hline
Docker Configuration and Patch Management Failures & Improper configuration or delayed security patches expose the system to vulnerabilities & CNs & Adherence to OWASP guidelines for secure Docker configuration \\
\hline
\end{tabular}
}
\vspace{-10pt}
\end{table*}

For CNs, the attached storage typically refers to the local storage of the physical or virtual machine where the containers are deployed. The local manager or user can access each experiment's global and local models to evaluate or deploy them, along with TensorBoard files to track their training process and performance.

In PS, the most recent global and local CN model versions are stored and organized by experiment ID.

Logging is implemented and log files are maintained in their respective experiment folders, ensuring comprehensive tracking mechanisms.

\section{Security Analysis}

Despite often being neglected in the existing literature, the security of FL systems is crucial given the project's critical nature and the sensitive data involved. A thorough security analysis from traditional distributed systems and FL perspectives is vital to address and mitigate security and privacy concerns, thereby enhancing the robustness and reliability of the solution. In particular, we focus on the security of the core components of the proposed platform: PS, CNs, and Communication Infrastructure. Table~\ref{tab:security_threats} outlines the threats, risks, and mitigation strategies of each component, demonstrating effective risk management through the design choices of the proposed platform.

\myparc{Parameter Server} Attacks can be executed by malicious users to the PS utilizing local updates to recover clients' local data due to conventional aggregation algorithms that are vulnerable to adversarial attacks \cite{inversion}. In the context of server-side security, the requirements primarily focus on monitoring vulnerabilities, authorization processes, access control, and the education of intern personnel. In addition, secure aggregation methods are required to prevent the server from leaking information and to detect anomalous updates from clients. From an infrastructure perspective, the network infrastructure at CERN already complies with these established requirements, further highlighting CERN’s appropriateness as serving as a public PS.

\myparc{Clinical clients} CNs are the most vulnerable component of the proposed platform, given their role in hosting sensitive data, operating on diverse IT infrastructures, training models with local data, and accessing the globally broadcasted model. The integrity of each node is crucial, as a compromised client could lead to data corruption or model updates through poisoning attacks, or exploit the global model for inference attacks \cite{inversion}. While the current design of the CN addresses many threats, further security enhancements are necessary. Key measures include implementing an access control system that limits machine access and adheres to PoLP, conducting regular penetration testing, maintaining patch management, and ensuring detailed logging and monitoring to improve threat detection and response.

\myparc{Communication infrastructure} The primary requirements for the communication infrastructure focus on client authentication, message integrity, and confidentiality to ensure secure channels that prevent data interception or tampering during frequent model update exchanges between nodes. The use of SSH tunnels, Kerberos, MQTT authentication primitives, and TLS encryption effectively meets these needs.

\section{Experimental Results with Stroke Data}
\label{sec:results}

The proposed platform was tested and evaluated using the publicly available Stroke Prediction Dataset \cite{dataset}, which predicts the likelihood of a stroke based on parameters such as gender, age, various diseases, and smoking status. Data was divided so that 20\% was reserved as a test set, while the remaining 80\% was distributed among three nodes, mimicking the target deployment and configuration. A five-fold cross-validation was performed and results are presented in terms of mean and standard deviation.

In the proposed tests, we compared the performance by considering three reference scenarios: \textbf{local training}, CNs use their local datasets to train local models independently without federated learning; \textbf{centralized training}, data is transferred to a central data center that supervises all learning stages centrally; and \textbf{federated training}, CNs are federation members and implement an assigned FL algorithm.

The model used in the experiments was a multi-layer perceptron with two layers, each with 512 units, using \texttt{tanh} activation and a dropout rate of 0.5. Settings between local, centralized, and federated experiments were matched as closely as possible to allow a fair comparison. Training was carried out for a maximum of 128 epochs or rounds, using the Adam optimizer with default parameters. A learning rate reducer on the plateau was set to trigger after 16 epochs or rounds, and early stopping was configured to 48 epochs or rounds. For federated aggregation, the four mentioned methods were tested.

Given that the dataset is highly imbalanced, F1-score and Area Under the Precision-Recall Curve (AUPRC) provide the most informative metrics for evaluation. The state-of-the-art results for this have an F1-score of around 30\%, providing a benchmark for comparison \cite{dataset}. Our results indicated that the local training scenario and FedDyn resulted in the lowest performance, while the centralized scenario achieved the highest performance. This centralized scenario represents the optimal technical solution from an ML perspective. However, it is not feasible in real-world healthcare applications, particularly TRUSTroke, due to data privacy concerns and regulations. FedAvg obtained the best federated result, representing a significant improvement over local training, highlighting the benefits of the federated approach and its successful implementation. SCAFFOLD and FedProx also provided an improvement over the local training. The performance of FedAvg over other aggreagation algorithms can be attributed to the relatively homogeneous distribution of the data between the simulated nodes, which favored the standard weighted averaging approach. In contrast, the other aggregation approaches were designed to address issues that arise in non-IDD data and might become crucial for real-world datasets even after employing a common data model across institutions.

The platform performed well throughout the experiments, validating the architecture, communication, and experiment tracking components. The significant improvements achieved through federated approaches, especially with FedAvg, demonstrate the feasibility and effectiveness of the FL architecture.

\begin{figure}
    \centering
    \includegraphics[width=1\linewidth]{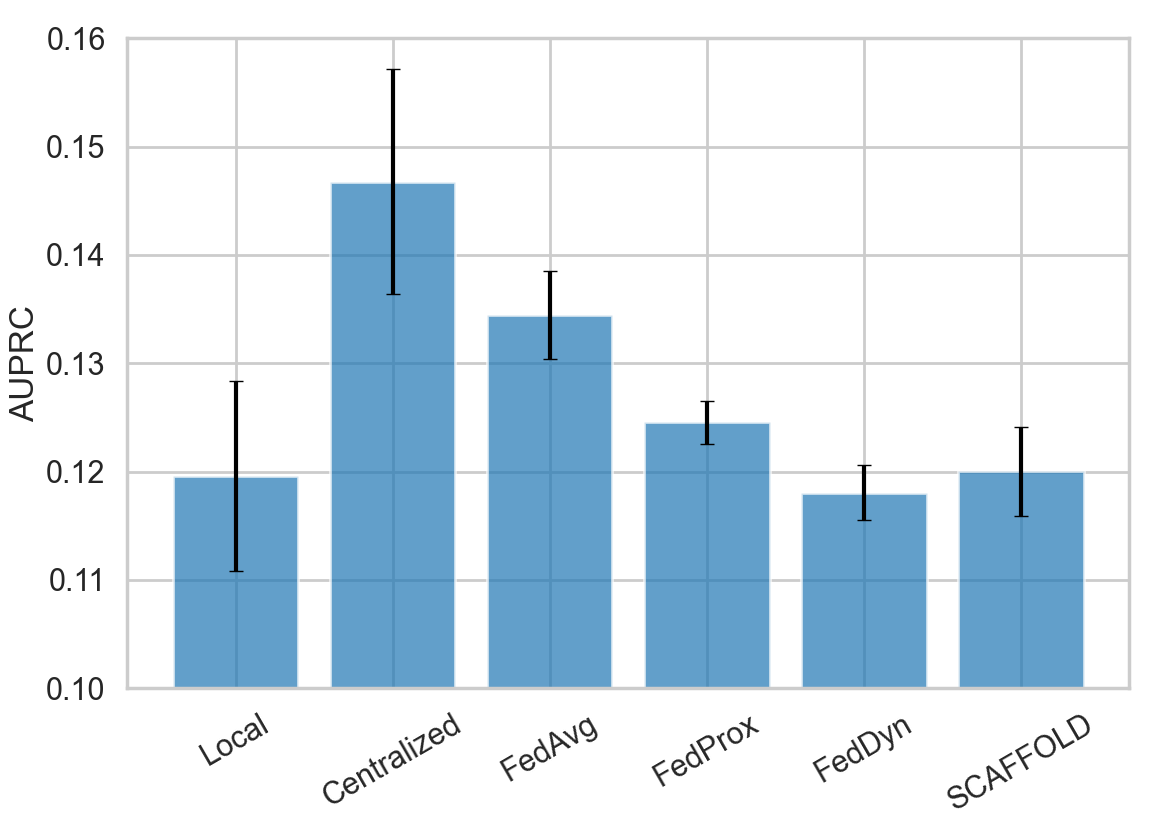}
    \vspace{-20pt}
    \caption{Comparative results using publicly available Stroke Dataset for local, centralized, and federated scenarios. Results are shown as the Area Under Precision-Recall Curve (AUPRC) mean and standard deviation from cross-validation.}
    \label{fig:enter-label}
\end{figure}

\begin{table}
    \caption{Comparative results using publicly available Stroke Dataset}
    \label{table:results}
    \begin{center}
    \vspace{-10pt}
    \resizebox{\linewidth}{!}{  
    \begin{tabular}{l|c|c|c|c}
    \hline
    \textbf{Method} & \textbf{Precision} & \textbf{Recall} & \textbf{F1 Score} & \textbf{AUPRC} \\
    \hline
    Local & 19.65 \scriptsize{$\pm$1.97} & 47.46 \scriptsize{$\pm$8.22} & 27.62 \scriptsize{$\pm$2.86} & 11.95 \scriptsize{$\pm$1.75} \\
    Centralized & 20.75 \scriptsize{$\pm$1.97} & 61.20 \scriptsize{$\pm$5.93} & 30.98 \scriptsize{$\pm$2.84} & \textbf{14.67} \scriptsize{$\pm$2.07} \\
    FedAvg & 17.09 \scriptsize{$\pm$0.91} & 70.00 \scriptsize{$\pm$2.00} & 27.47 \scriptsize{$\pm$1.26} & \textbf{13.44} \scriptsize{$\pm$0.80} \\
    FedProx & 15.64 \scriptsize{$\pm$0.46} & 66.80 \scriptsize{$\pm$1.09} & 25.34 \scriptsize{$\pm$0.65} & 12.45 \scriptsize{$\pm$0.39} \\
    FedDyn & 15.29 \scriptsize{$\pm$0.56} & 66.40 \scriptsize{$\pm$1.67} & 24.85 \scriptsize{$\pm$0.82} & 11.80 \scriptsize{$\pm$0.50} \\
    SCAFFOLD & 16.28 \scriptsize{$\pm$1.16} & 62.40 \scriptsize{$\pm$2.61} & 25.80 \scriptsize{$\pm$1.51} & 12.00 \scriptsize{$\pm$0.82} \\
    \hline
    \end{tabular}
    }
    \end{center}
\vspace{-10pt}
\end{table}

\section{Conclusions and future activities}
We have presented a robust FL platform tailored for healthcare applications and optimized for production environments. The platform consists of three modular applications that collectively facilitate secure, efficient, and privacy-preserving collaborative model training across multiple clinical sites.

The current implementation of the platform has been tested using a publicly available stroke dataset. These tests demonstrate that our FL platform operates correctly and offers significant improvements in model performance compared to local training while maintaining data privacy. This readiness for deployment is a crucial step toward integrating the platform into real-world hospital environment.

Security considerations have been thoroughly addressed, ensuring the platform's resilience against potential threats. The security measures implemented include secure data transmission, robust client authentication and authorization mechanisms, and the isolation of sensitive data within CNs.

Future work will focus on improving aggregation algorithms' security and performance, model security, and generally increasing the platform's reliability. Development efforts will integrate differential privacy techniques and model outlier detection modules to reinforce defenses against adversarial and non-adversarial attacks. Enhancements in authentication and access control will aim to deepen integration with CERN's existing systems, and improvements in logging will include centralized solutions to streamline monitoring and debugging processes. Furthermore, expanding the platform to support the training of distributed decision tree models will increase its applicability across common healthcare scenarios.

This platform represents a significant advancement in the application of FL to healthcare, specifically for the management of stroke patients. It enables secure, collaborative, and effective AI model training across multiple clinical sites.
\label{sec:conclusions}

\bibliographystyle{ieeetr}
\bibliography{bib}

\end{document}